\documentclass[12pt]{article}

\newcommand{\dpa}{\partial}

\begin{document}

\hspace*{0.29\textwidth} 
 \underline{\bf http://xxx.lanl.gov/e-print/physics/0205035}\\

\begin{center}
\textbf{\Large
Damping of electromagnetic waves in low-collision electron-ion plasmas}
\vspace{3mm}

V.~N.~Soshnikov
\footnote{Krasnodarskaya str., 51-2-168, Moscow 109559, Russia.}
\vspace{1mm}

Plasma Physics Dept.,\\
All-Russian Institute of Scientific and Technical Information\\
of the Russian Academy of Sciences\\
(VINITI, Usievitcha 20, 125315 Moscow, Russia)
\end{center}
\vspace{-2mm}

\begin{abstract}
  Using previously developed method~\cite{bib-1,bib-2} 
  of two-dimensional Laplace transform 
  we obtain the characteristic equations $k(\omega)$
  for electromagnetic waves in low-collision fully ionized plasma
  of a plane geometry.
  We apply here a new,
  different from the one used in~\cite{bib-1,bib-2},
  iteration procedure of taking into account 
  the Coulomb collisions. 
  The kinematical waves are collisionally damping 
  in the same extent as electromagnetic waves.
  Despite the different from~\cite{bib-2} appearance 
  of the dispersion (poles) equation,
  the obtained decrements for fast and slow wave modes 
  coincide with results obtained in~\cite{bib-2},
  if one neglects the terms of higher orders in $v_x^2/c^2$,
  ($v_x$ and $c$ are electron and light velocities).
  We point out how one can determine mutually dependent boundary conditions
  allowing to eliminate simultaneously 
  both the backward and kinematical waves 
  for transversal as well as for longitudinal oscillations.
\end{abstract}

PACS numbers: 52.25 Dg; 52.35 Fp.

Key words: {\em plasma oscillations; plasma waves;
                Landau damping; Coulomb collisions; collision damping;
                Vlasov equations; kinematical waves; plasma echo}.

\section{Introduction}

  Propagation of electromagnetic waves in low-collision 
fully ionized plasma
is described by asymptotic solution 
of the coupled kinetic and Maxwell equations.
The trivial fact is known:
an exponential solution $\exp(ikx - i\omega t)$, 
proposed by L.~Landau in 1946 
(in the simplest case of a plane geometry)
with complex $\omega$, 
is not a solution of either the Vlasov equations 
(for longitudinal plasma waves), 
nor the equations for transversal waves.
Nevertheless in the available literature 
one usually admits 
that namely Landau solution is true 
but the above mentioned equations must be correspondingly corrected 
by additional terms 
according to Landau rules of passing around poles 
in calculation of logarithmically divergent integrals 
appearing at substitution of the solution into the primary equations. 

    The proposed method of two-dimensional Laplace transformation
combined with Vlasov prescription of calculating divergent integrals 
in the sense of principal value 
allows one to obtain very simply asymptotical solutions 
of the original equations.
  
    In this work we briefly describe both the techniques 
and the results of the proposed in~\cite{bib-2}
new iteration procedure.
Following to this method one replaces Laplace image 
$Q_{p_1p_2}(\vec{v}_e)$
of the Coulomb collision term $Q(\vec{v}_e,x,t)$
by the term
\begin{equation}
 \label{eq-1}
   f_{p_1p_2}^{(1)}(\vec{v}_e)
    \left[\frac{Q_{p_1p_2}^{o}(\vec{v}_e)}
               {f_{p_1p_2}^{o}(\vec{v}_e)}
    \right]
   \,,
\end{equation}
where $f_{p_1p_2}^{(1)}(\vec{v}_e)$ is 
Laplace image of the perturbation of electron distribution function,
$Q_{p_1p_2}^{o}(\vec{v}_e)$ is Laplace image of the collision term,
calculated in the null-iteration approximation
using Laplace image $f_{p_1p_2}^{o}$
of the collisionless approximation of $f_{p_1p_2}^{(1)}$
\begin{equation}
 \label{eq-2}
  f_{p_1p_2}^{o}(\vec{v}_e)
  = \frac{|e|}{m_e}
    \cdot \frac{\dpa f_0^{(e)}(\vec{v})}{\dpa v_z} 
    \cdot \frac{E_{p_1p_2}}{p_1+v_xp_2}
  + \frac{v_xf_{p_1}^{(e)}(\vec{v})}{p_1+v_xp_2}\, .
\end{equation}
In our boundary problem we can suppose $f_{p_2}^{(e)}=0$.

  The method of subtraction of unphysical backward field waves
suggested in~\cite{bib-2} 
at non-zero $f_{p_1}^{(e)}$ and $f_{p_2}^{(e)}$
does not yet define single-valued dependence
of $f_{p_1}^{(e)}$ and $f_{p_2}^{(e)}$ 
on $\vec{v}_e$.
As we said before,
the boundary conditions are not independent 
and the given boundary electrical field $E(0,t)$
defines the boundary function $f_{1}^{(e)}(\vec{v},0,t)$.
Such an interrelation of these quantities 
can be determined through the natural condition 
of absence of kinematical waves. 
In the case $E(0,t)=E_0\cos(\omega t)$, $p_1=\pm i\omega$,
the general expression for $f_{p_1}^{(e)}(\vec{v})$
is 
\begin{equation}
 \label{eq-3}
   f_{p_1}^{(e)}(\vec{v})
   =
    \frac{a(\vec{v})}{p_1+i\omega}
   + 
    \frac{a^{*}(\vec{v})}{p_1-i\omega}
   \,,
\end{equation}
where symbol $^{*}$ means complex conjugation.

By equating amplitudes of the kinematical waves to zero
one obtains linear integral equations for determination
of $a(\vec{v})$ and $a^{*}(\vec{v})$:
\begin{equation}
 \label{eq-4}
  \left[\frac{|e|}{m_e}
   \frac{\dpa f^{(e)}_0}{\dpa v_z}
    E_{p_1p_2}
     \right]_{\begin{array}{l}
            p_1=\pm i\omega\\
            p_2=\mp p_1/v_x
            \end{array}}
  + v_x f^{(e)}_{p_1}(\vec{v})
  = 0
\end{equation}
where $E_{p_1p_2}$ (see Eq.(37) in~\cite{bib-2})
contains integrals of the type
\begin{equation}
 \int
  \frac{f^{(e)}_{p_1} u_x u_z d^3\vec{u}}
       {p_1+u_xp_2}\, .
 \label{eq-5}
\end{equation}
These equations define uniquely the dependence
$a(\vec{v})$ on $\vec{v}$.
But determined in this way $f_{p_1}^{(e)}(\vec{v})$
can not be used to eliminate unphysical backward field waves
in $E(x,t)$, as was supposed in~\cite{bib-2}.
To this end one must use 
the boundary condition for $F_{p_1}$
(that is Laplace transform of $\dpa E(x,t)/\dpa x|_{x=0}$).

In this way function $f_{p_1}^{(e)}$ has the form
\begin{equation}
 \label{eq-6}
  f_{p_1}^{(e)}(\vec{v}_e) 
  \sim 
  \frac{\dpa f_{0}^{(e)}}{\dpa v_z} 
   \eta(\vec{v})\, ,
\end{equation}
where $\eta(\vec{v})$ is some complicated function
of $\vec{v}$.
Assuming that factor $\dpa f_{0}^{(e)}/\dpa v_z$
is the main in the dependence of 
$f_{p_1}^{(e)}(\vec{v}_e)$ on $\vec{v}$
we can use Eq.(\ref{eq-6}) with
replacement $\eta(\vec{v})\to \eta(\bar{\vec{v}})$ 
for rough estimates.
Then, in the expression for collision term,
$f_{p_1}^{(e)}$ in Eq.(\ref{eq-2}) can be approximately omitted 
(both terms in Eq.(\ref{eq-2}) have the same structure in $\vec{v}$).

Analogous considerations in the case of longitudinal waves 
lead to the determination of $f_{p_1}^{(e)}$,
but there are no other free boundary conditions 
to eliminate the backward waves in $E(x,t)$.
This fact leads to the inevitable conclusion 
that the normal boundary component of electrical field
$E(0,t)=E_0\cos(\omega t)$
is broken at the plasma boundary due to the surface charge.
This plasma boundary field can be found 
with the proportional changing of $f_{p_1}^{(e)}(\vec{v})$
and $E_0 \to E_0'$, $E_{p_1} \to E_{p_1}'$
in linear equation of the type (\ref{eq-4})
without changing field amplitude $E_0$ in $E_{p_1}$
in equation for $E_{p_1p_2}$  (see Eq.(37) in~\cite{bib-2}).

\section{Collisional damping of electromagnetic waves}

  Characteristic equation which is an equation for double poles
$p_1$, $p_2$ of Laplace images of electrical field $E(x,t)$
and distribution function $f_{1}^{(e)}(\vec{v}_e,x,t)$
has been obtained in~\cite{bib-2}. 
For $E(0,t)\sim E_0\exp(\omega t)$ it has the following form
\begin{equation}
  G(p_1,p_2)
  = \left(p_1 - i\omega\right)
    \left[p_2^2 
       - \frac{p_1^2}{c^2}
       + \frac{\omega_L^2p_1}{c^2}
          \int \frac{v_z d^3\vec{v}}{p_1+v_xp_2} 
           \left(\frac{\dpa f^{(e)}_{0}}{\dpa v_z}
               - \frac{m_e}{|e|}
                 \frac{Q_{p_1p_2}^{o}(\vec{v})}{E_{p_1p_2}}
           \right)                                        
    \right] 
 = 0\, ,\label{eq-7}
\end{equation}
where $Q_{p_1p_2}^{o}(\vec{v}) \sim E_{p_1p_2}$
(see~\cite{bib-2}, Eqs.(27)-(29)).

    The integrals are defined according to Vlasov 
principal value prescription. 
The residue of Eq.(\ref{eq-7}) at pole $p_1=i\omega$
defines poles in $p_2$.

   In the case of procedure (1) the pole equation 
differs in form from the pole equation in~\cite{bib-2}:
\begin{equation}
  G(p_1,p_2)
  = \left(p_1 - i\omega\right)
    \left[p_2^2 
       - \frac{p_1^2}{c^2}
       + \frac{\omega_L^2p_1}{c^2}
          \int \frac{\dpa f^{(e)}_{0}}{\dpa v_z}
           \frac{v_z d^3\vec{v}}{p_1+v_xp_2-\Phi_{p_1p_2}(\vec{v})} 
    \right] 
 = 0\, ,\label{eq-8}
\end{equation}
where
\begin{equation}
 \label{eq-9}
  \Phi_{p_1p_2}(\vec{v})
  \equiv 
  \frac{Q_{p_1p_2}^{o}(\vec{v})}{f_1^0(\vec{v},p_1,p_2)}
\end{equation}
does not contain the value $E_{p_1p_2}$;
$f_1^0(\vec{v},p_1,p_2)$ is defined by Eq.(\ref{eq-2}).

   In analogy with~\cite{bib-2},
one uses approximation 
\begin{equation}
 \label{eq-10}
  \int_{-\infty}^{\infty}F(v_x)f_0^{(e)}(v_x)dv_x
  =
  \int_{0}^{\infty}\left[F(v_x)+F(-v_x)\right]f_0^{(e)}(v_x)dv_x
  \simeq
  \frac{F(v_{0x})+F(-v_{0x})}{2}\, ,  
\end{equation}
where coefficient $1/2$ appears owing to difference
in normalization of distribution functions 
taken in intervals $(-\infty, \infty)$ and $(0,\infty)$
and
\begin{equation}
 \label{eq-10a}
  v_{0x}\equiv\sqrt{\bar{v_x^2}}\, .
\end{equation}
Then one obtains
\begin{equation}
 \label{eq-11}
  \int \frac{\dpa f^{(e)}_{0}}{\dpa v_z}
   \frac{v_z d^3\vec{v}}{p_1+v_xp_2-\Phi_{p_1p_2}(\vec{v})} 
 \simeq
  \int_{-\infty}^{\infty} dv_y
  \int_{-\infty}^{\infty} 
   \Theta(v_{0x},v_y,v_z)
    \frac{\dpa f^{(e)}_{0}(v_{0x},v_y,v_z)}{\dpa v_z} v_z dv_z
\end{equation}
where
$$
  \Theta(v_{0x},v_y,v_z)
  \equiv
   \frac{\left[p_1-\frac{1}{2}\Phi_{p_1p_2}^{+}(v_{0x},v_y,v_z)\right]}
        {p_1^2-v_{0x}^2p_2^2
       - p_1\Phi_{p_1p_2}^{+}(v_{0x},v_y,v_z)
       + v_{0x}p_2\Phi_{p_1p_2}^{-}(v_{0x},v_y,v_z)
        }
$$
and
$$
  \Phi_{p_1p_2}^{\pm}(v_{0x},v_y,v_z)
   \equiv
    \Phi_{p_1p_2}(v_{0x},v_y,v_z) \pm \Phi_{p_1p_2}(-v_{0x},v_y,v_z)\, .
$$
The replacement $\Phi_{p_1p_2}(v_x,v_y,v_z)\to\Phi_{p_1p_2}(v_{0x},v_y,v_z)$,
$v_x^2 \to v_{0x}^2 \simeq kT/m_e$ 
is made after taking derivatives in $v_x$, $v_y$, and $v_z$ 
in the differential operator $Q_{p_1p_2}^{o}(\vec{v})$; 
integrals in $dv_y$ and $dv_z$ can be approximately
estimated by their mean values:
\begin{equation}
 \label{eq-12}
  \int_{-\infty}^{\infty} dv_y
   \int_{-\infty}^{\infty} 
    \Theta(v_{0x},v_y,v_z)
     \frac{\dpa f^{(e)}_{0}(v_{0x},v_y,v_z)}{\dpa v_z} v_z dv_z
   \simeq  N_{y} N_{z}
    \Theta(v_{0x},v_{0y},v_{0z})
\end{equation}
with evident normalization constants $N_y$ and $N_z$.

   After elementary transformations one obtains dispersion
(poles) equation in the form
\begin{equation}
 \label{eq-13}
  \left[\left(p_2^2 - \frac{p_1^2}{c^2}\right)
         \left(p_1^2 - v_{0x}^2p_2^2 - p_1a + v_{0x}p_2b\right)
       - \frac{\omega_L^2}{c^2}
          p_1\left(p_1 - \frac{a}{2}\right)
  \right] 
 = 0\, ,    
\end{equation}
where
\begin{eqnarray}
 \label{eq-14}
  a &\equiv&
  \Phi_{p_1p_2}^{+}(v_{0x},v_{0y},v_{0z})
  \ =\
  \frac{4\lambda}{\left(3v_{0x}^2\right)^{3/2}}
   \frac{p_1^4-2p_1^2v_{0x}^2p_2^2+5v_{0x}^4p_2^4}
        {\left(p_1^2-v_{0x}^2p_2^2\right)^2}
\\ \label{eq-15}
  b &\equiv&
  \Phi_{p_1p_2}^{-}(v_{0x},v_{0y},v_{0z})
  \ =\
  \frac{8\lambda}{\left(3v_{0x}^2\right)^{3/2}}
   \frac{p_1^3v_{0x}p_2-3p_1v_{0x}^3p_2^3}
        {\left(p_1^2-v_{0x}^2p_2^2\right)^2}
\\ \label{eq-16}
  \lambda &\equiv& \frac{2\pi e^4Ln_i}{m_e^2}\, ;\qquad
  p_1\ =\ i\omega\, ;\qquad
  v_{0x} \simeq v_{0y} \simeq v_{0z} \simeq \frac{kT_e}{m_e}\, ,
\end{eqnarray}
and further as
\begin{eqnarray}
  \left[\left(c^2p_2^2 - p_1^2\right)
         \left(p_1^2 - v_{0x}^2p_2^2\right)^3
      - \omega_L^2p_1^2\left(p_1^2 - v_{0x}^2p_2^2\right)^2
  \right] &&\nonumber\\
- \frac{4\lambda p_1}{\left(3v_{0x}^2\right)^{3/2}}
   \left(c^2p_2^2-p_1^2\right)
   \left[p_1^4 - 4p_1^2v_{0x}^2p_2^2 + 11v_{0x}^4p_2^4
   \right] &&\nonumber\\
+ \frac{2\lambda p_1}{\left(3v_{0x}^2\right)^{3/2}}
   \omega_L^2
   \left[p_1^4 - 2p_1^2v_{0x}^2p_2^2 + 5v_{0x}^4p_2^4
   \right] &=& 0\, . \label{eq-17}
\end{eqnarray}
This equation is an analogue of the characteristic equation (30)
in~\cite{bib-2}.

  For both electron modes
\begin{eqnarray}
 \label{eq-18}
  p_2^{(1)} 
  &=& \pm\frac{i\omega}{c}\sqrt{1-\frac{\omega_L^2}{\omega^2}}
    + \delta^{(1)}\, ,\\
 \label{eq-19}
  p_2^{(2)} 
  &=& \pm\frac{i\omega}{v_{0x}}\left(1+\frac{v_{0x}^2\omega_L^2}{2c^2\omega^2}\right)
    + \delta^{(2)}\, ,
\end{eqnarray}
we obtain from Eq.(\ref{eq-17}) at $|\delta^{(1,2)}|\ll 1$ and
neglecting terms with higher orders in $v_{0x}^2/c^2$ 
in Eqs.(\ref{eq-14}), (\ref{eq-15}):
\begin{eqnarray}
 \label{eq-20}
 \delta^{(1)}
 &=& \pm
     \frac{2\pi e^4 n_i L \omega_L^2}
          {3\sqrt{3}v_{0x} m_e kT_e c \omega^2
          \sqrt{1-\omega_L^2/\omega^2}}
 \, ;\\
 \label{eq-21}
  \delta^{(2)}
 &=& \pm 
     \left(\frac{\pi e^4 n_i L \omega^2}
                {3\sqrt{3}v_{0x}^4m_e kT_e}
     \right)^{1/3}\ .
\end{eqnarray}
that coincides with corresponding expressions for $\delta^{(1,2)}$
in~\cite{bib-2},
in spite of differences of characteristic equations.

    Let us emphasize here the sharp increase 
of the dissipative collisional absorption 
of electromagnetic waves 
proportional to $1/\sqrt{1-\omega_L^2/\omega^2}$ 
at $\omega\to\omega_L+0$
with the dominating collisionless non-dissipative reflective evanescence
of waves at $\omega<\omega_L$.

\section{Conclusions}

    Coincidence of collisional damping decrements 
for the two variants of iteration process 
is an evidence of the proposed calculation method correctness.
The approximation (\ref{eq-1}) is more preferable
than one used in~\cite{bib-1,bib-2}
since it has more evident physical sense.
The difference of both iteration procedures appears 
at large values of $v_{0x}^2/c^2$,
however in this case there is also growing contribution 
of relativistic corrections to the original equations.

    The requirements of absence of both unphysical 
backward (divergent at $x\to\infty$) field waves 
and kinematical waves smearing electron distribution function 
$f_1^{(e)}(\vec{v},x,t)$ in $v_x$
lead at the given boundary field $E(0,t)$
to the determination 
of the boundary distribution function $f_1^{(e)}(\vec{v},0,t)$
and the solution $f_1^{(e)}(\vec{v},x,t)$ 
with single-valued dependence on $\vec{v}$, $x$, and $t$.
\vspace{5mm}

\textbf{Acknowledgements}
  The author is thankful to Dr.~A.~P.~Bakulev 
for his criticism and assistance 
in preparing the paper in \LaTeX\ style.


\begin{thebibliography}{9}
\bibitem{bib-1}
 Soshnikov~V.~N.,
  "Damping of plasma-electron oscillations and waves 
   in low-collision electron-ion  plasmas",
  physics/0105040 (http://xxx.lanl.gov/e-print)
\bibitem{bib-2}
 Soshnikov~V.~N.,
  "Damping of transversal plasma-electron oscillations and waves
   in low-collision electron-ion  plasmas",
  physics/0111014 (http://xxx.lanl.gov/e-print)
\end{thebibliography}
\end{document}